# Transforming Wikipedia into an Ontology-based Information Retrieval Search Engine for Local Experts using a Third-Party Taxonomy

**Gregory Grefenstette, Karima Rafes**
Inria Saclay/TAO and BorderCloud, Rue Noetzlin - Bât 660
91190 Gif sur Yvette, France
{gregory.grefenstette,karima.rafes}@inria.fr, karima.rafes@bordercloud.com

**Abstract**

Wikipedia is widely used for finding general information about a wide variety of topicss. Its vocation is not to provide local information. For example, it provides plot, cast, and production information about a given movie, but not showing times in your local movie theatre. Here we describe how we can connect local information to Wikipedia, without altering its content. The case study we present involves finding local scientific experts. Using a third-party taxonomy, independent from Wikipedia's category hierarchy, we index information connected to our local experts, present in their activity reports, and we re-index Wikipedia content using the same taxonomy. The connections between Wikipedia pages and local expert reports are stored in a relational database, accessible through as public SPARQL endpoint. A Wikipedia gadget (or plugin) activated by the interested user, accesses the endpoint as each Wikipedia page is accessed. An additional tab on the Wikipedia page allows the user to open up a list of teams of local experts associated with the subject matter in the Wikipedia page. The technique, though presented here as a way to identify local experts, is generic, in that any third party taxonomy, can be used in this to connect Wikipedia to any non-Wikipedia data source.

**Keywords:** ontology, pivot ontology, Wikipedia, local search

## 1. Introduction

Wikipedia is a large multilingual crowd-sourced encyclopedia, covering millions of topics. Its goal is to provide a "compendium of knowledge" incorporating "elements of general and specialized encyclopedias, almanacs, and gazetteers."[1] Possessing over 5 million pages in the English version, Wikipedia involves two principal parties: hundreds of thousands of human editors who contribute new information and edit existing pages, and anonymous internet users who use Wikipedia as a resource for finding facts involving general knowledge. To help end users find information, Wikipedia contributors can also add general categories "to group together pages on similar subjects."[2] The categories added to Wikipedia is a graph and can contain cycles (Zesch and Gurevych, 2007) but it can be transformed into a taxonomy Ponzetto and Navigli (2009).

But one taxonomy is not good for all purposes (McDermott, 1999, Veltman, 2001), and different domains produce different taxonomies covering some of the same topics. For example, the Association for Computing Machinery publishes every four years or so its taxonomy (Santos and Rodrigues, 2009) that is used to annotate computer science articles for many conferences. This taxonomy or other available taxonomies are difficult to merge with Wikipedia category set "because the concepts of source taxonomies are in different granularity, different structure, ambiguous, and partly incompatible" (Amini, *et al.*, 2015). Though merging taxonomies is difficult (Swartout and Tate, 1999), taxonomies are useful for providing a hierarchic, faceted view on data (Hearst, 2006), and because they limit the conceptual space to the principal terms that interest the group for which the taxonomy was created.

The problem we address here is how to combine the

(i) domain-directed usefulness of a taxonomy, created for one group, with the

(ii) familiarity of use and search that Wikipedia offers, without altering the generality of Wikipedia's content, with a

(iii) local source of data independent from Wikipedia.

The use case we describe involves finding local experts for mathematical or computer science problem, the domain areas of the ACM classification, using Wikipedia as a search engine. Demartini (2007) proposed using Wikipedia to find experts by extracting expertise from Wikipedia describing people, or from Wiki editors that edited page corresponding to a given topic. West et al. (2012) tried to characterize what makes Wikipedia editors be considered as experts in a given field. Our approach differs from this previous work since it allows us to connect Wikipedia content and subject matter to people who do not appear in Wikipedia either as subjects or editors.

## 2. Finding Experts with Wikipedia

In large organizations, such as multinational corporations, universities, and even large research centers, it can be difficult to know who is an expert about a given subject. A common response to this problem is to create an ontology of expertise and manually or automatically assign, to experts, labels from this ontology. Beyond the cost or effort needed to produce the ontology, this solution creates an additional problem. Once such a knowledge base of experts exists, the searcher still has to internalize the ontology labels and their meaning in order to find the expert. The difficulty the user faces explains why some expert knowledge bases are found useful for one division in a large organization but be useless for another division which does not share the same terminology or perspective (Hahn & Subrami, 2000). We propose a method for finding experts that exploits a pivot ontology, hidden from the user, and which allows the searcher to browse Wikipedia, a resource that he or she is probably familiar with, in order to find his or her local expert.

[1] en.wikipedia.org/wiki/Wikipedia:Wikipedia_is_an_encyclopedia
[2] en.wikipedia.org/wiki/Help:Category

We suppose that the user of our system is familiar with Wikipedia, but would like to find some local experts to solve a given problem. We also suppose that the user is able to find the Wikipedia page concerning the problem they have in mind, i.e., the user knows how to navigate in the semantic space implicit in Wikipedia.

Since Wikipedia contains information about historical figures and entities of general and not necessarily local interest, we need to connect our set of local experts to the pages of Wikipedia. It would run counter to the philosophy of Wikipedia to alter its content to cover local events and local unhistorical information. One solution, then, would be to copy all of Wikipedia into a local wiki, and then we could modify the subject pages as we wished to include names of local experts, laboratories, companies that are concerned with every page of interest. Although this would solve the problem of connecting Wikipedia pages to local experts, this solution has a few drawbacks: (i) users would no longer be inside the "real Wikipedia" but in a copy, which would have to be kept up to date regularly, and accessed with a different URL, (ii) since Wikipedia contains over 4 million pages in its English version, one would have to make as many decisions as to which pages to modify.

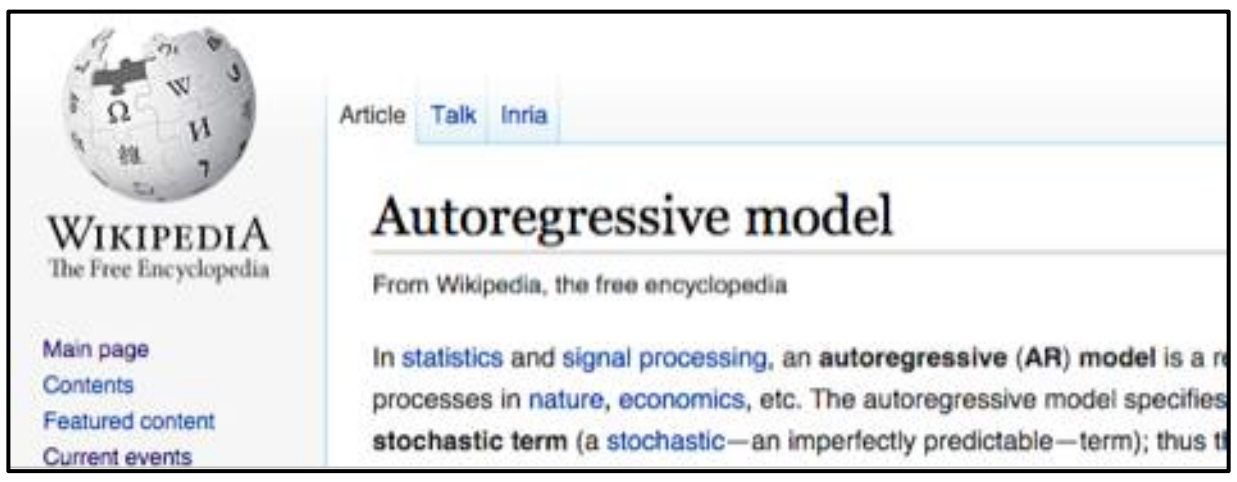

Figure 1 An extra tab appears while browsing Wikipedia once the expert finder module is activated by a logged-in user. Here the tab is labeled 'Inria' since we are searching for experts inside the Inria Institute

We have developed an automated technique that uses the same URL for Wikipedia, and which automatically establishes the connection between local expert and Wikipedia pages, using the ACM subject classification as a pivot ontology that is used to index the local experts and Wikipedia pages into the same semantic space, and a Wikipedia gadget (plugin) that adds a discrete tab to Wikipedia pages. To find a local expert for a given problem, the user searches the subject of interest in Wikipedia. The extra tab on each Wikipedia page can be clicked to reveal a list of experts concerning the some topic mentioned in the page.

## 2.1 Example

Before explaining the mechanisms and ontological resources involved, let us see an example. In this example, our local experts are any of the research teams in the French nation-wide computer science public research institute, Inria[3]. The Inria Institute employs 3600 scientists spread over 200 research teams, each specializing in some branch of computer science and mathematics. In our example, finding an expert will mean finding a team who is competent to answer questions about a given subject.

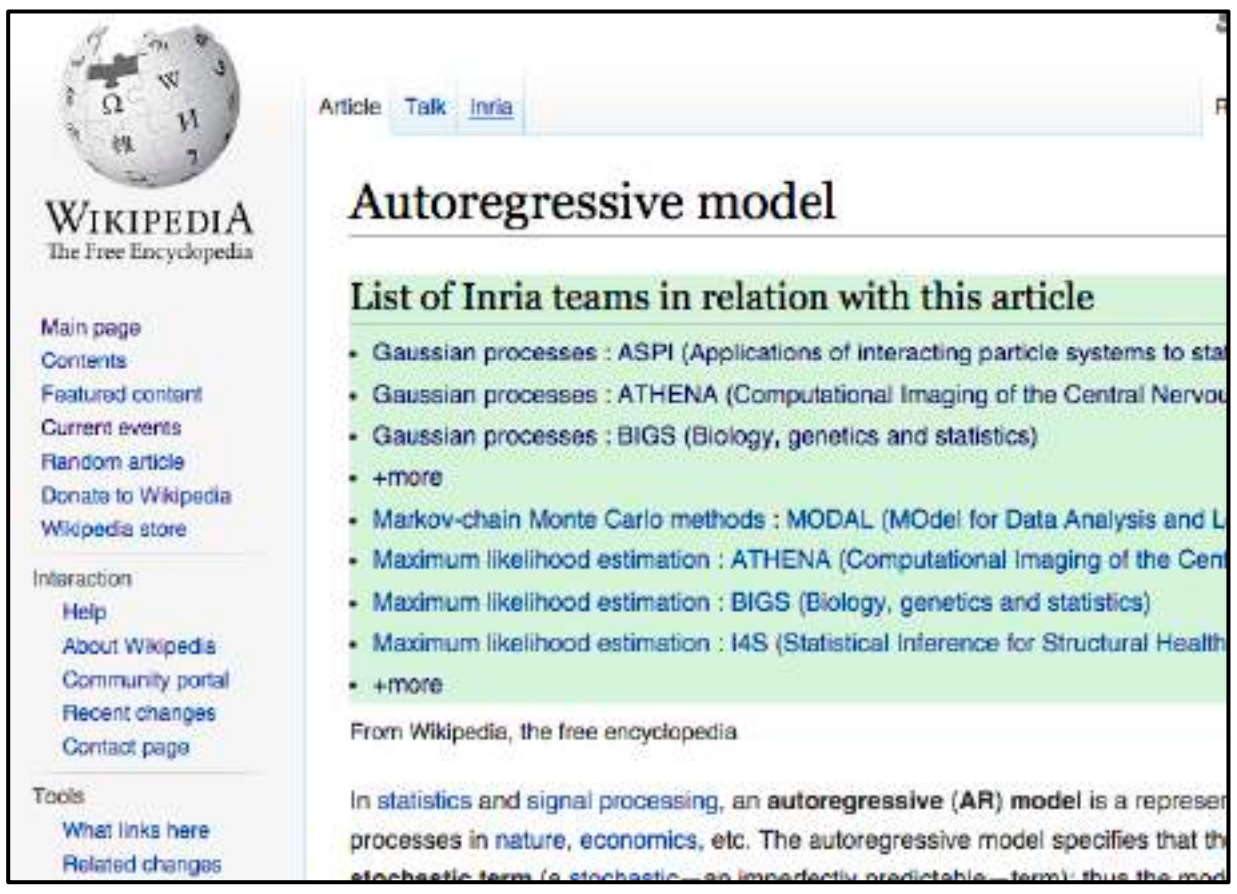

Figure 2. Clicking on the tab expands a box that shows the Inria teams that are associated with subjects on the Wikipedia page. For example, one subject mentioned on the page is "Gaussian processes" and 3 Inria teams that work in this domain are listed ASPI, ATHENA, and BIGS, with their expanded team names. Clicking on "Gaussian processes" goes to an ACM 2012 ontology page for this concept. Clicking on a team name goes to a web page from their 2014 annual report where a project involves "Gaussian processes". On this page, the user can find team members who are experts in the area. (See Figures 3 and 4)

In our solution, when someone is looking for an expert in a given subject inside the Inria institute, an additional tab appears to the identified user on the Wikipedia interface[4]. In Figure 1, the tab appears with the label "Inria" to the right of the "Article" and "Talk" tabs above the article title. Clicking on the tab expands a box listing the ACM subjects found in the articles and the Inria research teams treating those subjects, as seen in Figure 2.

[3] https://en.wikipedia.org/wiki/Inria

[4] This tab appears when the user is logged in, and has the resource module for expert finding, this tab appears while the user browses http://en.wikipedia.org. Wiki resource modules are additions that anyone can develop and activate. See further explanations at https://www.mediawiki.org/wiki/ResourceLoader/Developing_with_ResourceLoader

Both subjects and teams (see Figure 2) are linked to pages outside Wikipedia: if the user clicks on the subject name (e.g. “Maximum Likelihood Method”) in the expanded box, they are sent to the corresponding page in the ACM classification; if they click on the Inria team name (e.g. ATHENA), they are sent to the page in the team’s annual report which mentions the subject.

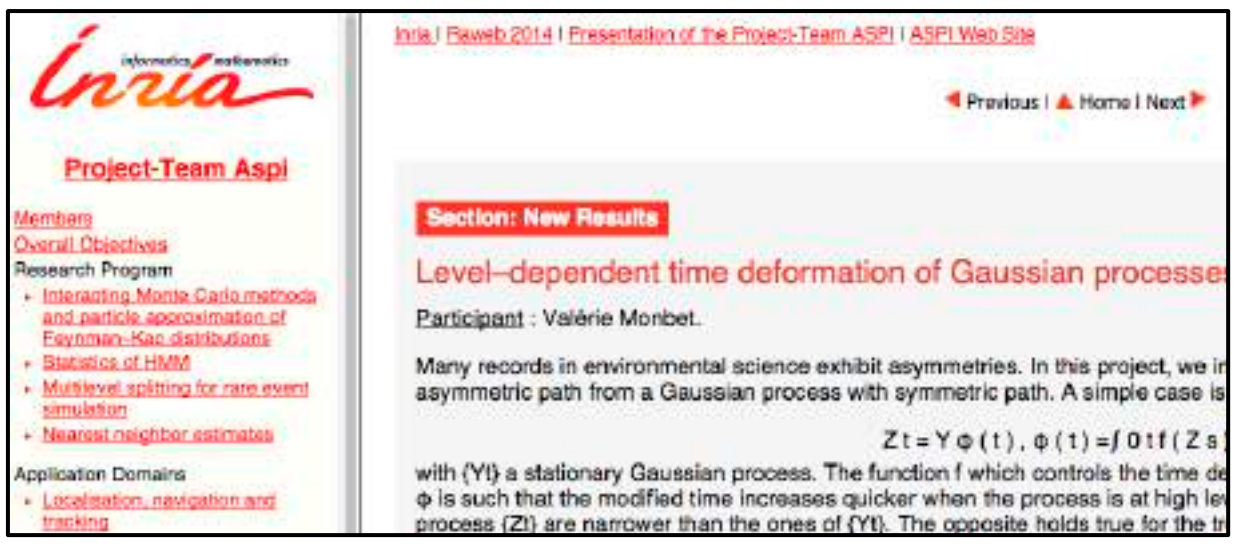

Figure 3. The Inria team annual report page found by following the link in the expert finding module. There the user sees that “Valerie Moribet” is involved in a project using Gaussian processes, and would probably be a good expert contact for finding out about Auto-regressive models.

Thus, Wikipedia has become a search engine with the user browsing towards their query (here “Autoregressive Models”, with the pull down expert box corresponding the search engine results page, leading to outside content. The user can find the Wikipedia article closest to his or her concern, and use the expert finding tab to find local experts who know about the subjects on the page.

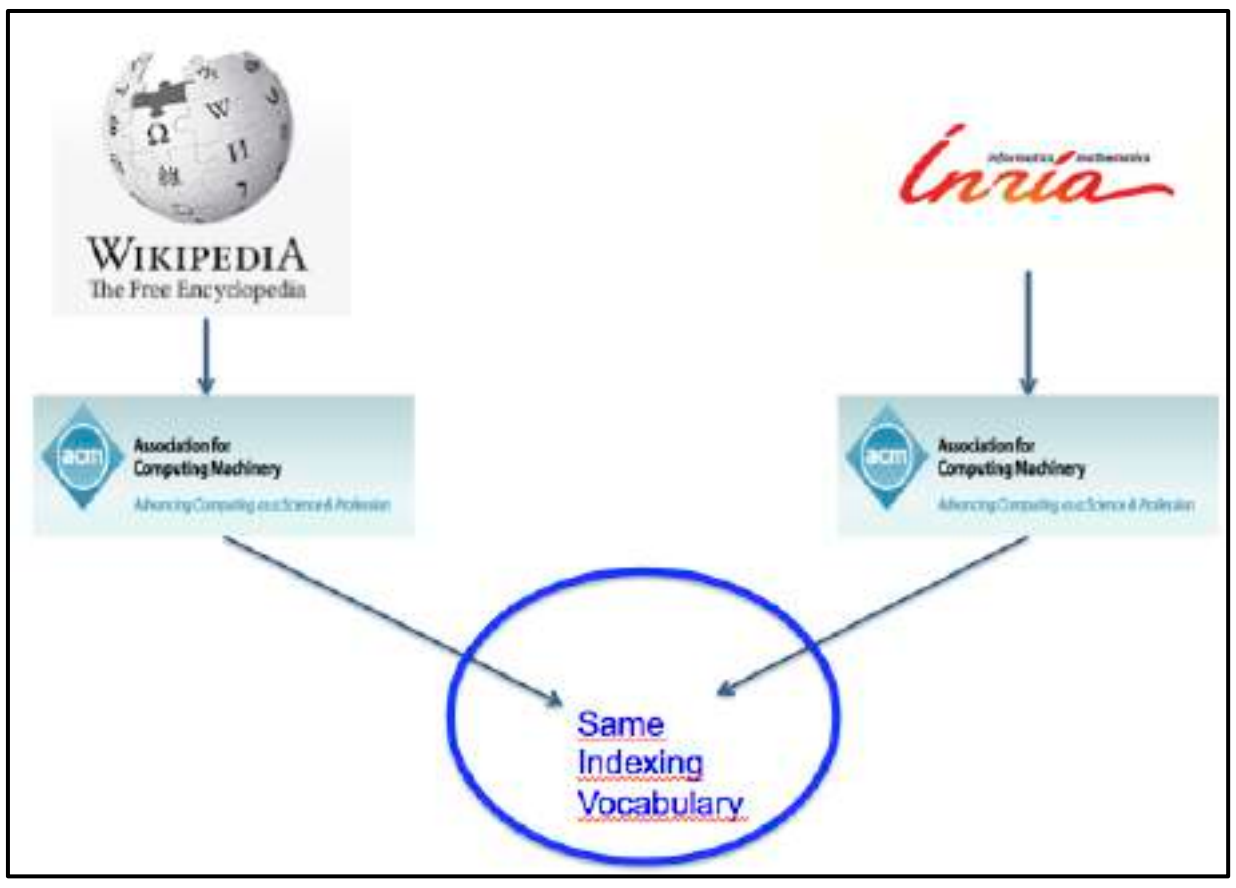

Figure 4. The ACM 2012 is the latest version of a computer science ontology created by the Association for Computing Machinery. "Gaussian processes" is a hyponym of “Theory of Computation”. This specific concepts provides a link between the Wikipedia web page on “Autoregressive models” and the Inria team experts, but the searcher need not understand the ACM hierarchy nor know its contents in order to make the connection.

This seems to us a natural and intuitive method for finding experts that obviates the need for learning the underlying, pivot ontology by which the experts are connected to the topic page. Even if the connecting ontology terms are explicitly displayed in the results, the user need not ever use them in an explicit query.

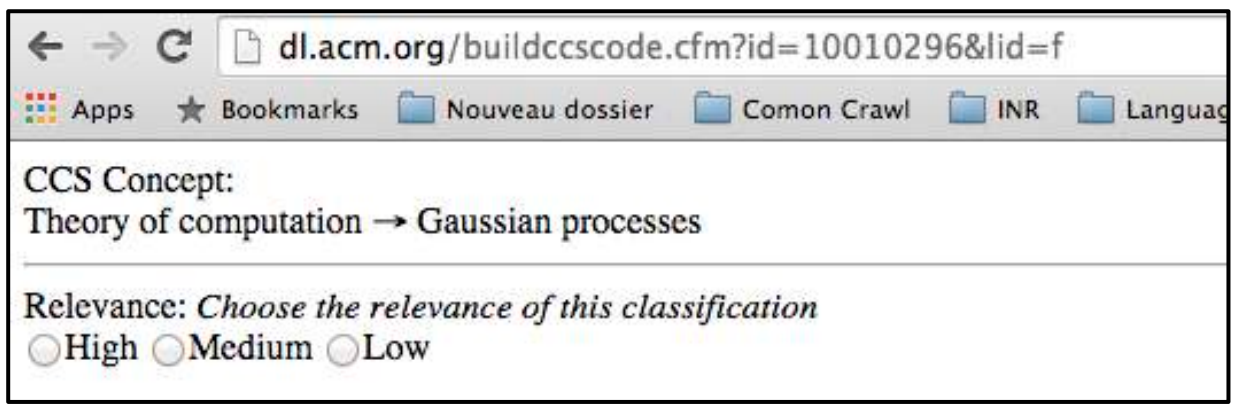

Figure 5. The content of all Wikipedia pages, and all information associated with the local expert teams are indexed using the same controlled vocabulary, here the ACM classification scheme.

## 2.2 Language Resources and Processing

Both Wikipedia page content and the expert profiles are mapped into the same ontology categories, which provide a pivot, or link between them. In our implemented example, we used the ACM 2012 Classification schema[5] as the shared ontological space. Here is an entry in this ontology:

<skos:Concept rdf:about="#10010296" xml:lang="en">
<skos:prefLabel xml:lang="en">Gaussian processes
</skos:prefLabel>
<skos:altLabel xml:lang="en">Gaussian process
</skos:altLabel>
<skos:inScheme
rdf:resource="http://totem.semedica.com/taxonomy/The ACM Computing Classification System (CCS)"/>
<skos:broader rdf:resource="#10010075"/>

This entry gives a synonym for “Gaussian processes”, an internal ACM code for this concept (10010296), and a link to a hypernym (10010075), “Theory of computation”. This SKOS taxonomy, augmented by any Wikipedia redirects as additional synonyms, was converted into a natural language processing program that associates the internal ACM code to raw English text (delivered as resource with this paper).

Wikipedia text was extracted from a recent dump.[6] The text was extracted from each article, tokenized, lowercased, and transformed into sentences. If any preferred or alternative term from the ACM classification scheme was found, then that ACM code was associated to the Wikipedia article[7]. Any ACM concept appearing in more than the adhoc threshold of 10,000 articles was eliminated as being too general[8].

The source data for expert profiles in our implementation are the public web pages of Inria teams 2014 activity reports[9]. Each page was downloaded, boilerplate removed, text extracted, tokenized lowercased and split into sentences, as with the Wikipedia text.

[5] http://www.acm.org/about/class/class/2012

[6] http://dumps.wikimedia.org/biwiki/latest

[7] The programs for turning English text into sentences, and for recognizing ACM codes in text can be found athttp://pages.saclay.inria.fr/gregory.grefenstette/LRECpack.gz

[8] This threshold eliminates concepts ‘Women’ found under ‘Gender’.

In all 3123 Inria web pages were associated 129,499 Wikipedia articles were tagged one or more of with 1049 different ACM codes.

For example, the Wikipedia page "Artificial Intelligence" contains the ACM classification phrase "supervised learning". Many of the Inria research team annual report web pages also mention this topic, for example, the page uid70.html of the Inria team TAO. A link between the Wikipedia page "Artificial Intelligence", the term "supervised learning" and the TAO web page is created and stored in a publicly accessible SPARQL endpoint.

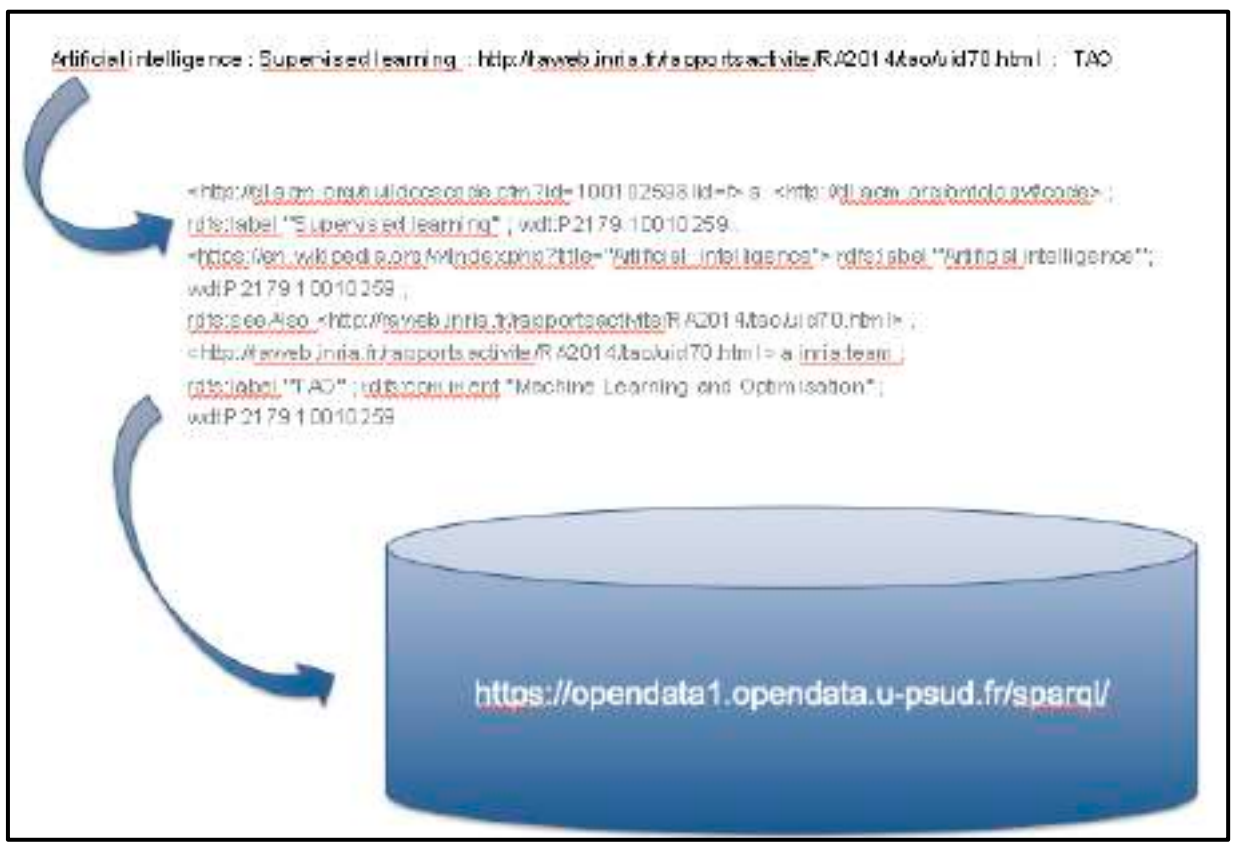

Figure 6. The link between each concept found in each Wikipedia page, and each outside data source, is rewritten as RDF and stored in a publicly accessible SPARQL database.

This SPARQL endpoint is accessed by the program "DisplayInriaTeam2.js" which is embedded in a Wikipedia gadget. In order to install this gadget in your Wikipedia;

1) visit wikipedia.org and register on the site to create an account
2) log in, and go to the following page: https://meta.wikimedia.org/wiki/Special:MyPage/global.js
3) create the page, and add the following line to it:

*mw.loader.load('//www.wikidata.org/w/index.php?title=User:ggrefen/DisplayInriaTeam2.js&action=raw&ctype=text/javascript');*

4) refresh your browser and visit a page with a computing topic
5) you can de-activate this gadget by adding two slashes (//) at the beginning of the line (before the word *mw.loader.load*).

Once you have followed the above steps, every time you log in, this gadget is activated and the "Inria" tab is inserted to every Wikipedia page, accessing the SPARQL data base when clicked on and producing the expanded box (see Figure 2) of the ACM concepts present on the page, and the Inria teams concerned with these concepts, providing the explicit links between the Wikipedia page content and the local external data.

## 2.3 Variations

We have currently implemented this search solution linking Wikipedia to Inria research team annual reports. Instead of using annual reports to create the "expert profiles", one could instead use the publications of researchers from a given team or research center. For example, each Inria team is obliged to publish their papers in the open access repository hal.inria.fr and the title of the papers, or their abstracts, or their full content could also be used to link Wikipedia page content to individual researchers. In this case, a different SPARQL endpoint could instantiated in this gadget, or a different gadget could be used for this local data. In place the ACM hierarchy as a pivot ontology[10], one could extract a taxonomy of Wikipedia categories and subcategories[11], or use another existing ontology, to index the Wikipedia content and the outside expert profiles. For example, one could use MeSH as the anchor ontology and publications of doctors at local hospital to transform Wikipedia into a search engine of specialists for medical problems.

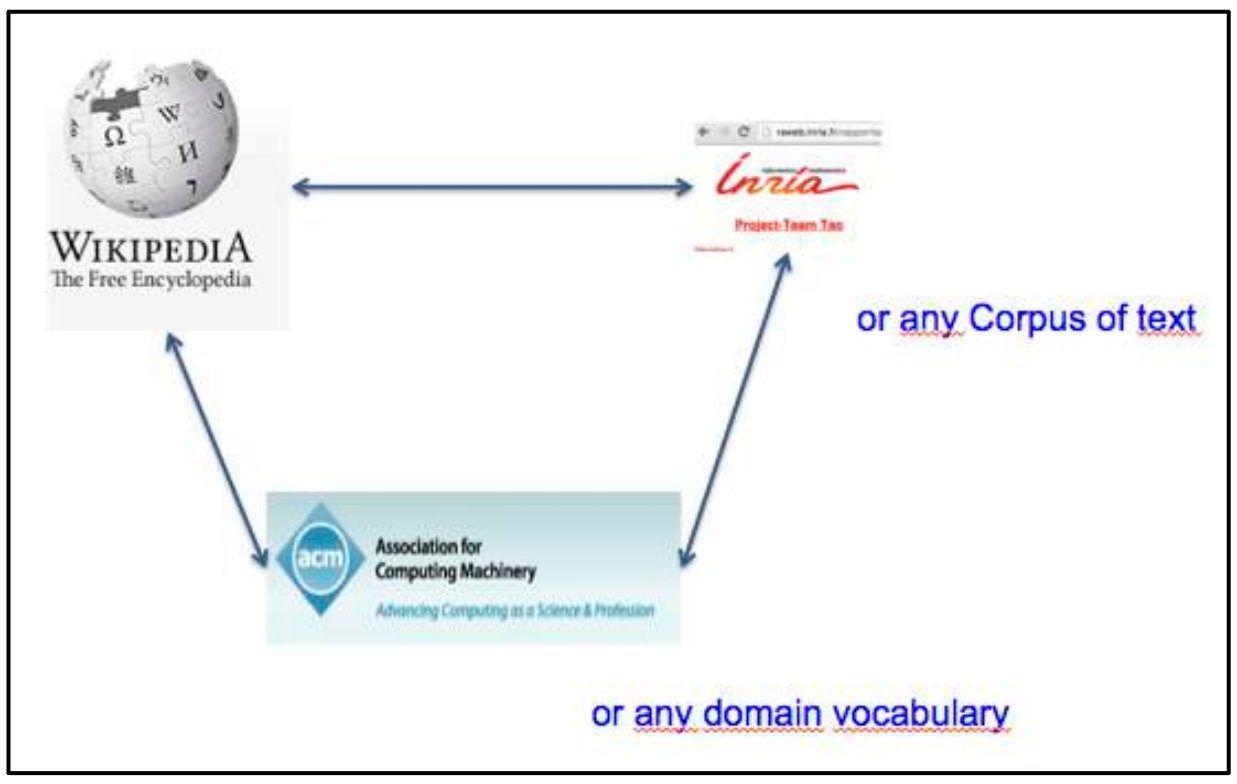

Figure 7. Our approach allows us to substitute and textual data source as the local expert repository, and any domain vocabulary as the pivot ontology. Thus Wikipedia, wihout altering its source text, can be used to search over any local data repository.

# 3. Conclusion

We have presented an Ontology-based information retrieval system for finding local experts using Wikipedia. It is constructed using a pivot ontology, indexing Wikipedia pages, and some textual representation of the experts (web pages, reports, or publications). The pivot ontology must be represented as a natural language processing resource, i.e. a resource that a program can be used to index to natural language text, that is applied to both Wikipedia pages and the textual representation of experts, moving them to the same space defined by the vocabulary of the resource. Once this mapping is done, and once the expert finding tab is opened, a Wikimedia resource loader dynamically makes the connection between the Wikipedia subject matter and the local experts by accessing a SPARQL endpoint, and constructs the list of local experts which is inserted into the displayed Wikipedia text. Neither the text of Wikipedia, not the expert profile text are permanently altered. An additional advantage of this system is that the user seeking an expert does not interact explicitly with the pivot ontology, but only with Wikipedia and the original textual representations of the experts. One plan to improve this system is to produce a more contextual markup, rather than the current string matching, by first categorizing the Wikipeida page as belonging to the subject of interest (in the case of Inria, this would be mathematics and computing).

[9] https://raweb.inria.fr/rapportsactivite/RA2014/index.html

[10] As a side effect of our work, Wikidata has decided to include the class system of the ACM in its hierarchy of concepts.

[11] Starting from, for example, https://en.wikipedia.org/wiki/Category:Computer_science

As the general public becomes more used to using Wikipedia to find information, they are able to find the best page that characterizes their need. With this gadget, Wikipedia is transformed into a local expert search engine[12].

### 3.1 Acknowledgments

This work is supported by the Center for Data Science, funded by the IDEX Paris-Saclay, ANR-11-IDEX-0003-02, and by an Advanced Researcher grant from Inria.

## 4. Bibliographical References

[12] 3 other Wikipedia gadgets for local data sets have also been developed using the LinkedWiki platform (Rafes & Germain, 2015):
https://io.datascience-paris-saclay.fr/appDisplayDatasets.php
https://io.datascience-paris-saclay.fr/appDisplayDevices.php
https://io.datascience-paris-saclay.fr/appDisplayScientists.php